\documentclass[a4paper,11pt]{article}
\usepackage{pos}
\usepackage{pos}
\usepackage{epsfig}
\usepackage{bm}

\usepackage{amssymb}
\usepackage{fontenc}
\usepackage{times}
\usepackage{graphicx}

\title{Chiral Symmetry Breaking in QED induced by an External Magnetic Field}

\author*[a]{D.~K.~Sinclair}
\author[b]{J.~B.~Kogut}

\affiliation[a]{HEP Division, Argonne National Laboratory, 
9700 South Cass Avenue, Lemont,Illinois 60439, USA}

\affiliation[b]{Department of Energy, Division of High Energy Physics,
Washington, DC 20585, USA \\
and\\
Department of Physics -- TQHN, University of Maryland, 82 Regents Drive,
College Park, MD 20742, USA}

\emailAdd{dks@anl.gov}
\emailAdd{jbkogut@umd.edu}

\abstract{We simulate Lattice QED in a constant and homogeneous external 
magnetic field using the Rational Hybrid Monte-Carlo (RHMC) algorithm
developed for Lattice QCD. Our current simulations are directed towards 
observing chiral symmetry breaking in the limit of zero electron bare mass
as predicted by approximate (Schwinger-Dyson) methods. Our earlier simulations
were performed on a $36^4$ lattice at the fine structure constant 
$\alpha=1/137$, close to its physical value, with `safe' electron masses
$m=0.1$ and $m=0.2$. At this $\alpha$, the dynamical electron mass produced 
by the external magnetic field, which is an order parameter for this chiral 
symmetry breaking, is predicted to be far too small to be measurable. Hence we
are now simulating at the larger $\alpha=1/5$, where the predicted dynamical 
electron mass at strong external magnetic fields accessable on the lattice is 
large enough to be measurable. However this requires electron masses down to 
$m=0.001$. Such a small $m$ requires lattices larger than $36^4$, but at 
magnetic fields large enough to produce measurable dynamical electron masses, 
$36$ is an adequate spatial extent for the lattice in the plane orthogonal to 
the magnetic field because the electrons preferentially occupy the lowest 
Landau level. We are therefore performing finite size analyses using 
$36^2 \times N_\parallel^2$ lattices with $N_\parallel \geq 36$. We measure 
the chiral condensate $\langle\bar{\psi}\psi\rangle$ as our order parameter for
chiral symmetry breaking, since it should remain finite as $m \rightarrow 0$ 
if chiral symmetry is broken by the magnetic field, but vanish otherwise. Our 
preliminary results strongly suggest that chiral symmetry {\it is} broken 
by the external magnetic field. In all our simulations, as well as measuring
other observables during these simulations, we are storing configurations at
regular intervals for further analysis. One such measurement planned for these
stored configurations is the determination of the effects that an external 
magnetic field has on the coulomb field of a charged particle placed in this 
magnetic field.
}

\FullConference{%
The 39th International Symposium on Lattice Field Theory,\\
8th-13th August, 2022,\\
Rheinische Friedrich-Wilhelms-Universität Bonn, Bonn, Germany
}

\begin{document}
\maketitle

\section{Introduction}

We study Lattice QED in external electromagnetic fields using methods 
developed for Lattice QCD. Since QED in background electric fields has a
complex action, because the vacuum is unstable against decays into electron-%
positron pairs -- the Sauter-Schwinger effect, \cite{Sauter,Schwinger:1951nm}
-- standard simulation methods which rely on importance sampling cannot be 
used. Because of this we start by considering Lattice QED in background 
magnetic fields where the action is real and bounded below. This enables us to
perform simulations using standard methods. We use the Rational Hybrid 
Monte-Carlo (RHMC) method of Clark and Kennedy \cite{Clark:2006fx} whose 
implementation we describe in the appendix.

Relativistic quantum mechanical studies of electrons in external 
electromagnetic fields and the modified actions this produces for those fields,
by Sauter, Euler and Heisenberg \cite{Sauter,Heisenberg} and formalized by 
Schwinger \cite{Schwinger:1951nm}, are some of the earliest QED calculations. 
See Dunne \cite{Dunne:2012vv} for some of the cases where exact solutions are 
known.

Phenomena such as the Sauter-Schwinger effect only become significant when
the electric field $E \sim E_{cr}=m^2/e$ or larger and/or the magnetic field 
$B \sim B_{cr}=m^2/e$ or larger. Interest in extending such studies to full 
QED including non-perturbative effects have been revived by planned 
experiments colliding electron beams with intense beams of light from petawatt
lasers \cite{Fedeli:2020fwt,Meuren:2021qyv}, where the electromagnetic fields 
are of this magnitude, or larger at LBNL, SLAC and possibly ELI. In addition 
it has been realized that compact astronomical X- and $\gamma$-ray sources are 
probably neutron-stars with magnetic fields of order $B_{cr}$ or larger 
(magnetars). See for example the review \cite{Harding:2006qn}. Finally it has 
been noted that beam-beam interactions in the next generation of 
electron/positron colliders could produce electromagnetic fields orders of 
magnitude larger than their critical values \cite{Yakimenko:2018kih}. Here, 
multi-electron-loop contributions become as, or more important than, single 
loop contributions and all conventional QED calculations break down.   

For our simulations of QED in an external magnetic field, we choose a magnetic
field which is constant over all space and time. For definiteness, we choose
a magnetic field oriented in the $z$ ($3$) direction. Classically the orbit
of a charged particle (electron) in such a magnetic field is a helix around
a fixed magnetic field line, whose projection on the $(x,y)$ ($(1,2)$) plane is
a circle, while the motion in the $z$ direction is free. Quantum mechanically
the motion in the $(x,y)$ plane is quantized into a set of levels whose 
transverse energies squared are evenly spaced with spacing $|2eB|$ -- the 
Landau levels\cite{Akhiezer}. The lowest level has a single helicity, while the
higher levels have both helicities. The the motion along the $z$ direction is 
free field so that the $z$ momenta have a continuous spectrum. Including QED 
means adding a dynamical photon field. The electron field feels both the 
external and the dynamical photon fields, while only the dynamical photon 
field has a kinetic term. For details of the lattice transcription of QED in 
this external field and its simulation, see the appendix. The most important 
feature is that at large $|eB|$ all electrons preferentially occupy the lowest
Landau level whose orbit has a finite extent (proportional to $1/\sqrt{|eB|}$,
leading to an effective dimensional reduction from $3+1$ dimensions to $1+1$ 
dimensions.

One of the most theoretically interesting non-perturbative effects predicted 
by truncated Schwinger-Dyson analyses of QED in such constant magnetic fields 
is that, in the limit $m \rightarrow 0$, chiral symmetry is broken by the
magnetic field leading to a dynamical electron mass $\propto\sqrt{|eB|}$  
\cite{Gusynin:1994xp,Gusynin:1995nb,Gusynin:1999pq,Gusynin:2000tv,
Leung:1996poh,Alexandre:2000nz,Alexandre:2001vu,Wang:2007sn} and a chiral 
condensate $\langle\bar{\psi}\psi\rangle\propto|eB|^{3/2}$ 
\cite{Shushpanov:1997sf,Lee:1997zj}. This non-perturbative effect is often 
referred to as `magnetic catalysis'. For a good review article with a more 
complete set of references see Miransky and Shovkovy \cite{Miransky:2015ava}. 
We note that in $3+1$ dimensions, for massless electrons in an external 
magnetic field without QED, i.e. without internal photons, chiral symmetry is 
unbroken for all $eB$. In fact, as $m \rightarrow 0$ the chiral condensate 
vanishes $\propto m \log(m)$, so QED is essential for the breaking of chiral 
symmetry in a magnetic field. This contrasts with the situation in $2+1$ dimensions where chiral symmetry is broken with a finite chiral condensate 
$\propto eB$ for massless electrons in an external magnetic field, even 
without QED. 

In our RHMC lattice QED simulations, we measure this chiral condensate, since 
as a local operator, it is easier to measure than the dynamical mass, which 
would require measuring the electron propagator itself. At physical 
$\alpha=e^2/(4\pi)\approx1/137$, the predicted dynamical mass is more than
30 orders of magnitude less than any value we could possibly measure. We
therefore perform simulations with a stronger $\alpha=1/5$, which appears to
be in the perturbative regime for $eB=0$, and for which the predicted 
dynamical electron mass and chiral condensate although small, should be
measurable for the $eB$ value we choose. Following our earlier simulations at
$\alpha=1/137$ we simulate on a $36^4$ lattice at $\alpha=1/5$ and with masses
in the range $0.001 \le m \le 0.2$. We choose 
$|eB|=2\pi\times100/36^2=0.4848...$, which is large but safely in the range
$|eB|<0.65$ required to keep discretization errors under control. For these
parameters a lattice of size $36$ in both the $x$ and $y$ directions is
considerably larger than the projection of the lowest Landau level on the
$(x,y)$ plane and therefore adequate. However, $36$ is too small a value in 
the $z$ and $t$ directions to accommodate the smallest masses, so a finite 
size scaling analysis is needed, however it is only necessary to increase the 
lattice sizes in the $z$ and $t$ directions. Preliminary results of such a 
finite size scaling analysis are presented in the next section, and strongly 
suggest that there is chiral symmetry breaking in the $m \rightarrow 0$ limit.
 
\section{Simulations and Results}

Here we discuss only the simulations and results from our simulations with
$\alpha=1/5$. For simulations and results with $\alpha=1/137$ and for `free'
electrons in an external magnetic field see the proceedings from our talk at
Lattice 2021 \cite{Sinclair:2021plv}.

We perform RHMC simulations with $\alpha=1/5$ aimed primarily at searching for
evidence for chiral symmetry breaking in the presence of a constant and uniform
magnetic field in the limit $m \rightarrow 0$. For this we measure the chiral
condensate $\langle\bar{\psi}\psi\rangle$, which should remain finite and non 
zero as $m \rightarrow 0$ if chiral symmetry is broken in this limit. 

\begin{figure}[htb]
\parbox{2.9in}{
\epsfxsize=2.9in
\epsffile{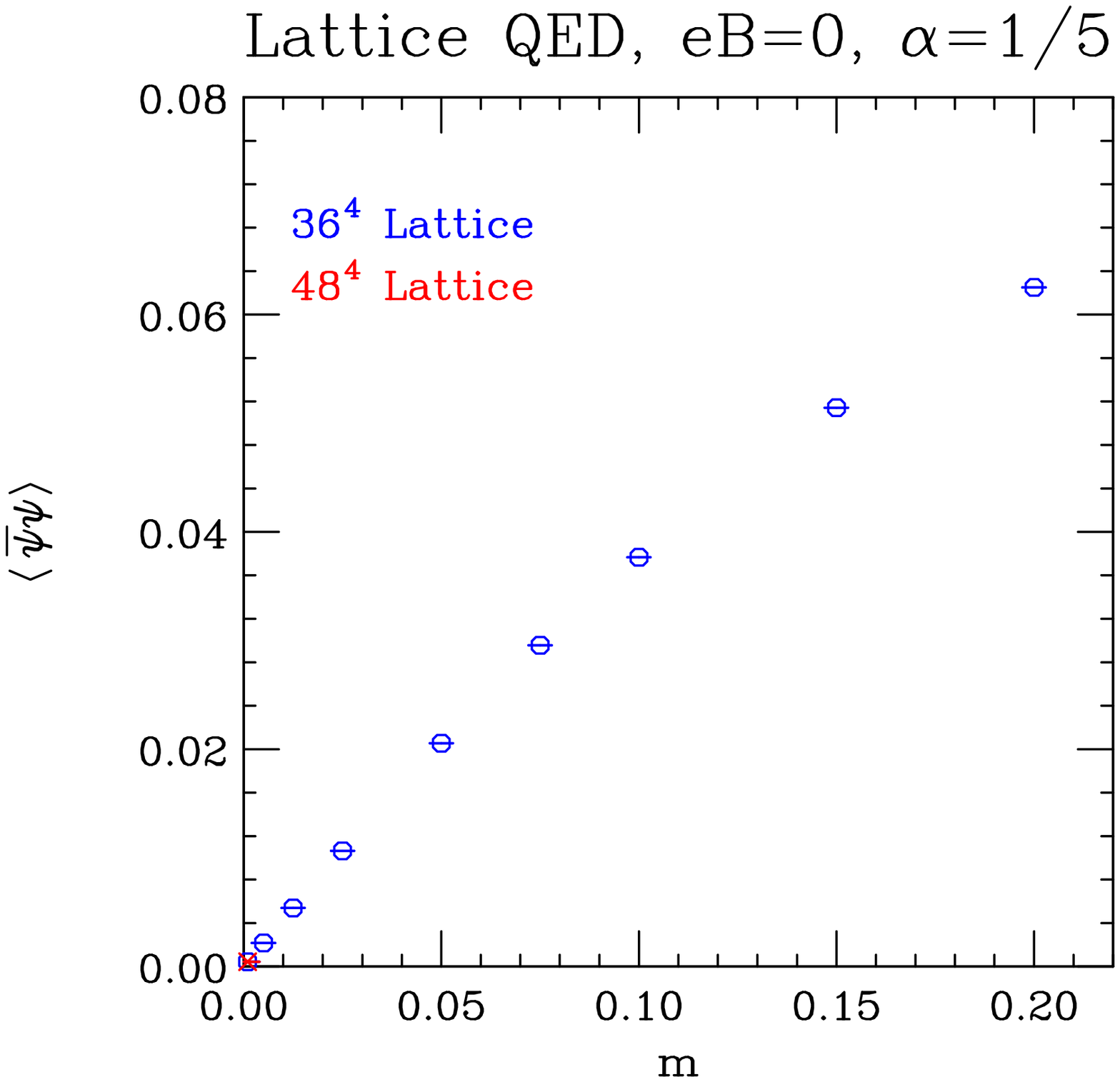}
}
\parbox{0.2in}{}
\parbox{2.9in}{
\epsfxsize=2.9in
\epsffile{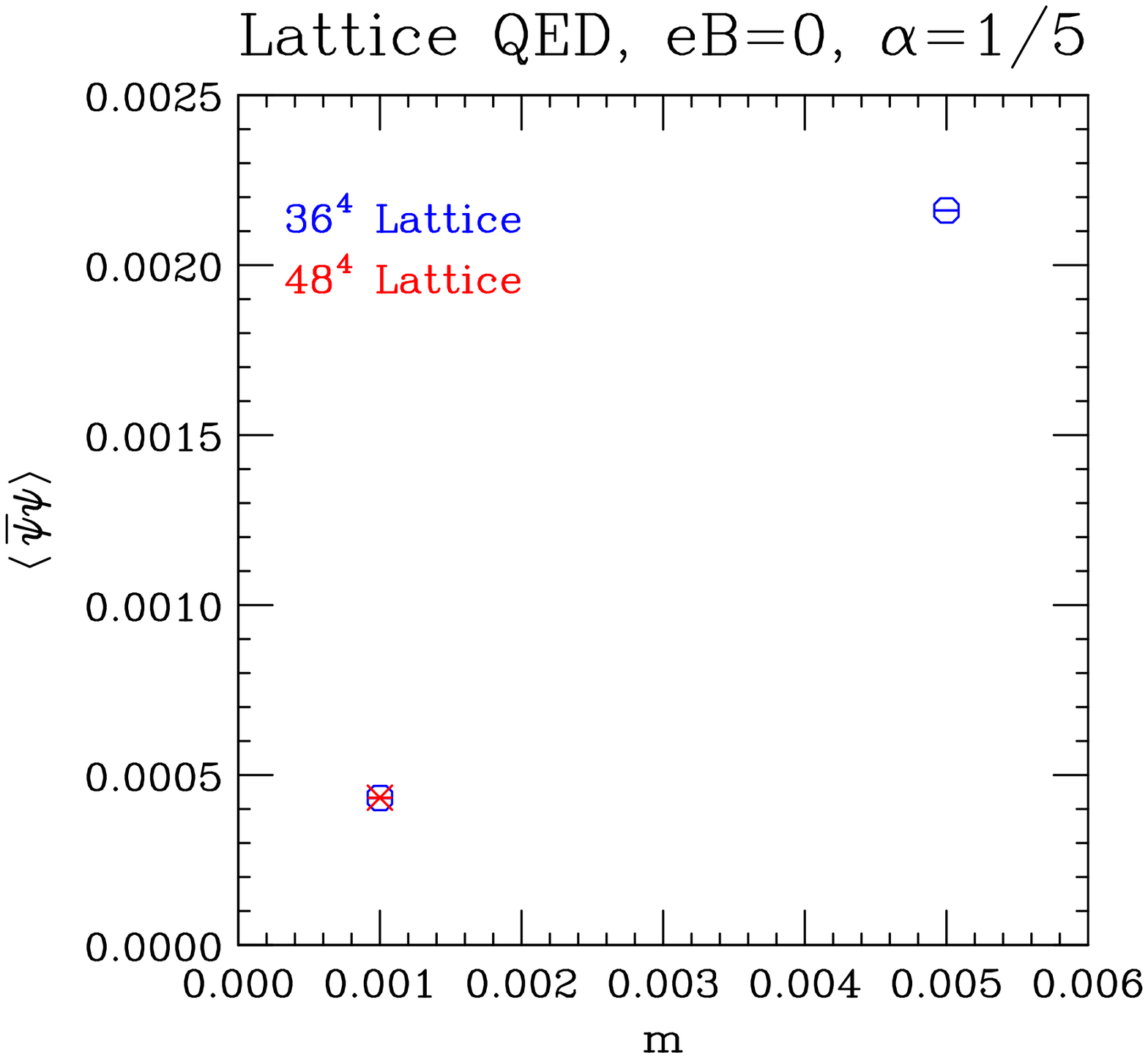}
}
\caption{$\langle\bar{\psi}\psi\rangle$ as a function of mass at $eB=0$,
showing lattice size dependence.}
\label{fig:pbp_b0}
\end{figure}

If chiral symmetry is unbroken in the $m=0$ limit, the chiral condensate is
dominated by the short distance (ultraviolet) regime and should vanish
proportional to $m$ possibly times some power of $\log(m^2)$ as 
$m \rightarrow 0$. In this case it should be insensitive to the size of the
lattice. Therefore we should be able to run with arbitrarily small masses
without observing finite lattice size effects. To test this we perform
$\alpha=1/5$ simulations with $0.001 < m < 0.2$ on a $36^4$ lattice at $eB=0$,
where chiral symmetry is believed to be unbroken at $m=0$. Note that the
normal requirement that $mN_\mu >> 1$ ($\mu=1,2,3,4$) to avoid finite size 
effects is not true for the lower part of this range. In 
figure~\ref{fig:pbp_b0} we plot the condensate as a function of mass and see 
that it does appear to be approaching zero for small $m$. We repeat these 
simulations on a $48^4$ lattice at the lowest mass $m=0.001$. The condensate on
the $36^4$ lattice is $4.3259(7) \times 10^{-4}$, while that on the $48^4$ 
lattice is $4.3294(7) \times 10^{-4}$, a mere less than $0.1\%$ difference. 
Such insensitivity to this finite size scaling analysis we take as evidence 
that the condensate is zero and chiral symmetry remains unbroken in the chiral
($m=0$) limit, at least to within the precision of our simulations. In fact, a
linear extrapolation to $m=0$ from the points at $m=0.005$ and $m=0.001$ which,
based on the curvature of this graph, should yield an upper estimate of the
value of the condensate at $m=0$, gives 
$\langle\bar{\psi}\psi\rangle \approx 10^{-6}$ which is only about 3 standard
deviations from zero, giving further evidence that chiral symmetry remains
unbroken as $m \rightarrow 0$.

We now turn to the case of large $eB$ and choose 
$|eB|=2\pi\times100/36^2=0.4848...$, near the upper end of the range of $|eB|$
values where discretization errors are small. Again we run our RHMC simulations
with $\alpha=1/5$ and $0.001 < m < 0.2$ on a $36^4$ lattice. As indicated 
above, the extent of the lattice in the $x$ and $y$ directions is large
enough to contain the lowest Landau levels. However the extent of the lattice
in the $z$ and $t$ directions is insufficient to prevent finite size effects.
If chiral symmetry is broken in the limit $m \rightarrow 0$ then this indicates
that there are modes with momenta of order $\sqrt{|eB|}$ or less which 
contribute to the condensate at small $m$. These modes will make a 
contribution of order $|eB|^{3/2}$ to the chiral condensate, keeping it 
non-zero as $m \rightarrow 0$ and making it sensitive to increases in the 
lattice extent in the $z$ and $t$ directions 
\cite{Shushpanov:1997sf,Lee:1997zj}. 
  
\begin{figure}[htb]
\parbox{2.9in}{
\epsfxsize=2.9in
\epsffile{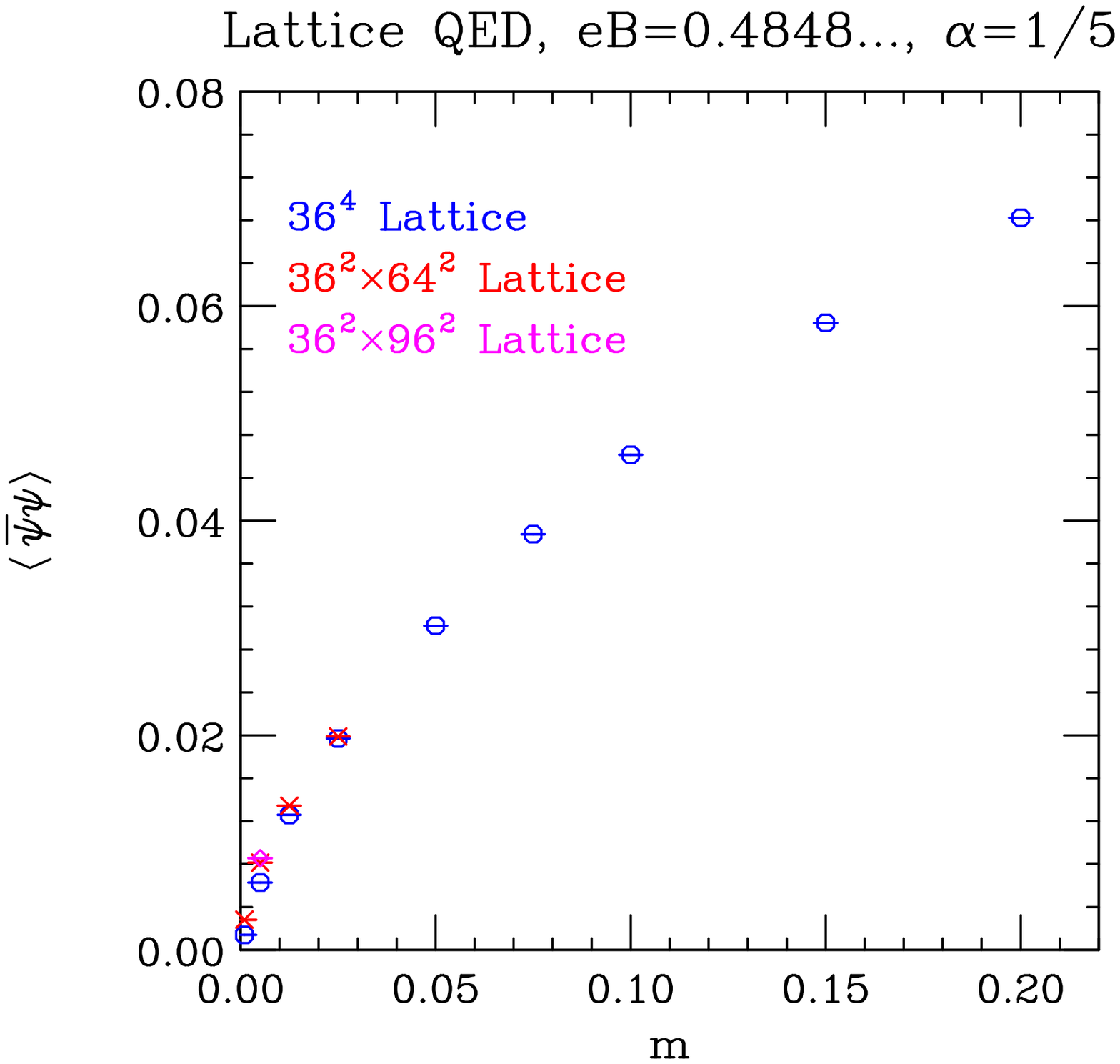}
}
\parbox{0.2in}{}
\parbox{2.9in}{
\epsfxsize=2.9in
\epsffile{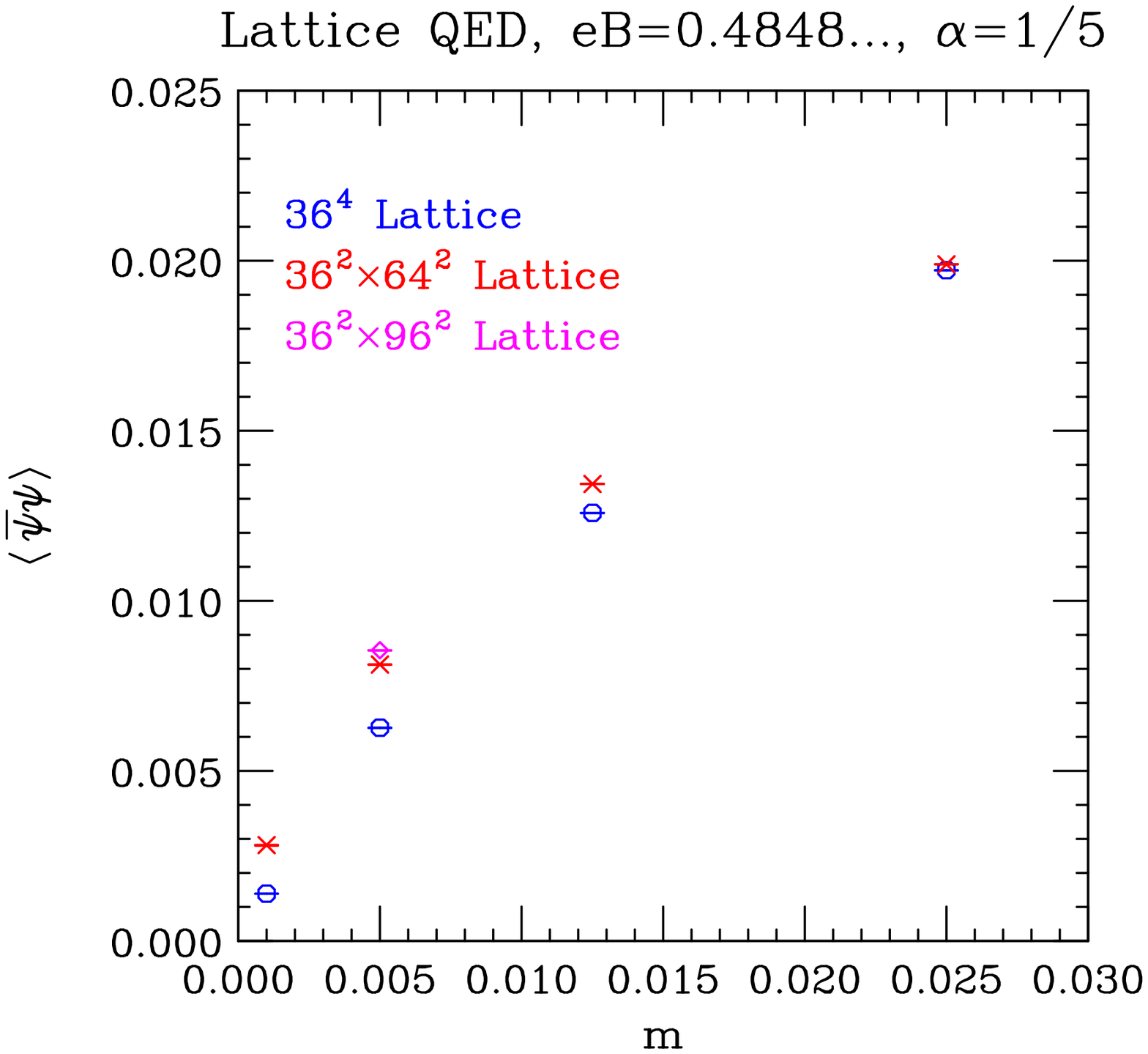}
}
\caption{$\langle\bar{\psi}\psi\rangle$ as a function of mass at 
$eB=2\pi\times 100/36^2=0.4848...$, showing dependence on lattice size in the 
$z$ and $t$ directions.}
\label{fig:pbp_b100}
\end{figure}

Since the condensate on the $36^4$ lattice appears to be headed towards zero
as $m \rightarrow 0$, a sign that chiral symmetry is broken in this limit
is that the chiral condensate at small enough $m$ should increase when the
lattice sizes in the $z$ and $t$ directions are increased. We have therefore
performed simulations with the same parameters on $36^2 \times 64^2$ lattices.
For $m=0.025$ the change in the chiral condensate in going from the $36^4$
lattice to the $36^2 \times 64^2$ lattice is very small, indicating that we 
need only simulate on the larger lattice at masses less than $0.025$. At 
$m=0.0125$ there is a small but significant increase in the condensate in going
to the larger lattice, while at $m=0.005$ the increase in the condensate in 
going to the larger lattice is relatively large. At m=0.001 the condensate on 
the larger lattice is approximately twice that on the smaller lattice. At 
$m=0.005$, we  have increased our lattice size even further to 
$36^2 \times 96^2$. While this leads to a further significant increase in the 
condensate, it is only by $\approx 5\%$, and so going to an even larger lattice
is unnecessary. Our next task is to increase the lattice size at $m=0.001$ to 
$36^2 \times 96^2$ and possibly go to a lattice with $z$ and $t$ extents of 
$128$, which should make the case for chiral symmetry breaking even more 
compelling and allow us to estimate the value of the chiral condensate at 
$m=0$. Figure~\ref{fig:pbp_b100} shows the mass dependence of the chiral 
condensate as a function of mass $m$ from these simulations at 
$|eB|=2\pi\times100/36^2$ on our chosen lattice sizes. Even with simulations on
only $36^4$ and $36^2 \times 64^2$ lattices at $m=0.001$, this graph does 
suggest that the condensate remains finite and non-zero as $m \rightarrow 0$.

\section{Summary, Discussion and Conclusions} 

We simulate lattice QED on lattice sizes $36^4$, $36^2 \times 64^2$, and
$36^2 \times 96^2$ with $\alpha=1/5$, in an external magnetic field 
$eB=2\pi\times100/36^2$ and masses in the range $0.001 \leq m \leq 0.2$.
The simulations on the $36^2 \times 96^2$ lattice at $m=0.001$ have yet to be 
performed. Preliminary results of this finite size scaling analysis show that 
for $m \leq 0.0125$, the chiral condensates increase as the lattice size is 
increased, which is evidence that the condensate remains finite in the chiral 
($m \rightarrow 0$) limit. The forthcoming larger lattice simulations at 
$m=0.001$ should confirm this and predict the value of this condensate at 
$m=0$. We also perform simulations with the same $\alpha$ and masses, but with
$eB=0$ on $36^4$ lattices and for $m=0.001$ on a $48^4$ lattice. Here the 
condensate shows no significant finite size effects, and the data strongly 
indicate that this condensate vanishes for $m \rightarrow 0$.

The size of the chiral condensate at $m=0$ appears to be larger than that 
predicted by Schwinger-Dyson methods. This might mean that $\alpha=1/5$ is
too large for the approximations made in the Schwinger-Dyson calculations or
the derivation of the condensate from the dynamical electron mass predicted
by these methods. It could also be due to the limitations of the lattice
methods at these parameters.

Because the size of the $(x,y)$ projection of lowest Landau levels for the 
chosen $eB$ are appreciably smaller than $36^2$, we are simulating with the
same parameters on lattices with dimensions $18^2$ in the $(x,y)$ plane for
comparison, in particular comparing the chiral condensates for small $m$.

We plan to perform further analyses on stored configurations. Of particular 
interest is the measurement of the distortion and screening of the coulomb
fields of charged particles in the presence of external magnetic fields
\cite{Shabad:2007xu,Shabad:2007zu,Sadooghi:2007ys,Machet:2010yg}

It is of interest to repeat our simulations in different external magnetic 
fields to test if the $m=0$ chiral condensate scales as $|eB|^{3/2}$. We should
also study the chiral behavior of QED extended to include multiple electron
`flavours'. If chiral symmetry is broken in the $m \rightarrow 0$ limit, this
includes the spontaneous breaking of chiral flavour symmetry, complete with
flavoured massless Goldstone bosons. These are allowed because, being 
uncharged, they do not feel the magnetic field and are thus not subject to the
dimensional reduction from $3+1$ to $1+1$ dimensions, and hence are massless 
$3+1$ dimensional excitations.

We will explore the possibility of designing an effective action which 
incorporates the assumption that only the lowest Landau level contributes,
but is otherwise fully $3+1$ dimensional.

We will explore the inclusion of external electric fields. Because these make
the action complex, we will need to resort to simulation methods such as
the complex Langevin equation (CLE). Because the non-compact gauge action 
describes a free field which is a collection of harmonic oscillators then, in 
the absence of fermions, the real Lagrangian is an attractive fixed point of
the CLE, rather than a repulsive fixed point as is the case for QCD and 
probably lattice QED with a compact action. Therefore one might hope that this
might remain true when fermions are included, possibly with a modified fermion
action, at least for weak coupling. This would allow the study of the 
Sauter-Schwinger effect in full QED in an external electric field using Lattice
QED simulations.

\acknowledgments

DKS's research is supported in part U.S. Department of Energy, Division of
High Energy Physics, under Contract No. DE-AC02-06CH11357. The high performance
computing was provided by the LCRC at Argonne National Laboratory on their
Bebop cluster. Access to Stampede-2 at TACC, Expanse at UCSD and Bridges-2 at
PSC was provided under an XSEDE allocation. Time on Cori at NERSC was provided
through an ERCAP allocation. DKS thanks G.~T.~Bodwin for inciteful discussions,
while JBK would like to thank V.~Yakimenko for discussions which helped inspire
this project, and I.~A.~Shovkovy for helpful discourse on magnetic catalysis.

\appendix\section{Lattice QED in an external magnetic field}

We simulate using the non-compact gauge action
$$
S(A) = \frac{\beta}{2}\sum_{n,\mu < \nu}[A_\nu(n+\hat{\mu})-A_\nu(n)
                                        -A_\mu(n+\hat{\nu})+A_\mu(n)]^2
$$
where $n$ is summed over the lattice sites, and $\mu$ and $\nu$ run from $1$ to
$4$ subject to the restriction $\mu < \nu$. $\beta=1/e^2$. The expectation 
value of an observable ${\cal O}(A)$ is then
$$
\langle{\cal O}\rangle = \frac{1}{Z}\int_{-\infty}^\infty \Pi_{n,\mu}
           dA_\mu(n) e^{-S(A)}[\det{\cal M}(A+A_{ext})]^{1/8}{\cal O}(A)
$$
where ${\cal M} = M^\dag M$, $A$ is the dynamic photon field and $A_{ext}$ is
the external photon field while
$$
M(A+A_{ext}) = \sum_\mu D_\mu(A+A_{ext})+m
$$
where the operator $D_\mu$ is defined by
\begin{eqnarray*}
[D_\mu(A+A_{ext})\psi](n) &=&
\frac{1}{2}\eta_\mu(n)\{e^{i(A_\mu(n)+A_{ext,\mu}(n))}\psi(n+\hat{\mu})\\
&-&e^{-i(A_\mu(n-\hat{\mu})+A_{ext,\mu}(n-\hat{\mu}))}\psi(n-\hat{\mu})\}
\end{eqnarray*}
and $\eta_\mu$ are the staggered phases.

We use the RHMC simulation method of Clark and Kennedy, using rational
approximations to  ${\cal M}^{-1/8}$ and ${\cal M}^{\pm 1/16}$. To account for
the range of normal modes of the non-compact gauge action, we randomly vary
the trajectory lengths over the range of periods of these modes
\cite{Hands:1992uv}. $A_{ext}$ are chosen in the symmetric gauge in the x-y
plane so that the magnetic fields from each plaquette are in the +z-direction
and have the value $eB$ modulo $2\pi$. This requires $eB=2\pi n/(n_1 n_2)$, 
where $n_1$ and $n_2$ are the lattice dimensions in the x and y directions, and
$n$ is an integer in the range $[0,n_1 n_2/2]$ \cite{Alexandre:2001pa}.

One of the observables we calculate is the electron contribution to the
effective gauge action per site $\frac{-1}{8V}{\rm trace}[\ln({\cal M})]$.
For this we use a rational approximation to $\ln$ following Kelisky and Rivlin
\cite{KR}, and a stochastic approximation to the trace.


\end{document}